\documentclass[conference]{IEEEtran}

\IEEEoverridecommandlockouts
\usepackage{bbm}
\usepackage{amsfonts}
\usepackage[dvips]{graphicx}
\usepackage{times}
\usepackage{cite}
\usepackage{amsmath}
\usepackage{array}
\usepackage{amssymb}

\usepackage{stfloats}
\usepackage{slashbox}
\usepackage{graphicx}
\usepackage{footnote}
\usepackage{booktabs}
\usepackage{array}
\usepackage{algorithm}
\usepackage{subeqnarray}
\usepackage{cases}
\usepackage{threeparttable}
\usepackage{color}
\usepackage{hyperref}
\usepackage{epstopdf}
\usepackage{algpseudocode}
\usepackage{bm}
\usepackage{multirow}
\usepackage[labelformat=simple]{subcaption}
\usepackage{adjustbox}
\makeatletter
\def\ps@headings{%
\def\@oddhead{\mbox{}\scriptsize\rightmark \hfil \thepage}%
\def\@evenhead{\scriptsize\thepage \hfil \leftmark\mbox{}}%
\def\@oddfoot{}%
\def\@evenfoot{}}
\makeatother 

\begin{document}

\title{Two-Way Passive Beamforming Design for RIS-Aided FDD Communication Systems}

\author{Bei Guo, Chenhao Sun and Meixia Tao\\
Department of Electronic Engineering, Shanghai Jiao Tong University, Shanghai, China\\
Email: \{guobei132, sjtusch, mxtao\}@sjtu.edu.cn
\thanks{This work is supported by the NSF of China under grant 61941106.}}
\maketitle

\vspace{-1.5cm}
\begin{abstract}
 Reconfigurable intelligent surfaces (RISs) are able to provide passive beamforming gain via low-cost reflecting elements and hence improve wireless link quality. This work considers two-way passive beamforming design in RIS-aided frequency division duplexing (FDD) systems where the RIS reflection coefficients are the same for downlink and uplink and should be optimized for both directions simultaneously. We formulate a joint optimization of the transmit/receive beamformers at the base station (BS) and the RIS reflection coefficients. The objective is to maximize the weighted sum of the downlink and uplink rates, where the weighting parameter is adjustable to obtain different achievable downlink-uplink rate pairs. We develop an efficient manifold optimization algorithm to obtain a stationary solution. For comparison, we also introduce two heuristic designs based on one-way optimization, namely, time-sharing and phase-averaging. Simulation results show that the proposed manifold-based two-way optimization design significantly enlarges the achievable downlink-uplink rate region compared with the two heuristic designs. It is also shown that phase-averaging is superior to time-sharing when the number of RIS elements is large. 
\end{abstract}
\section{Introduction}
 Reconfigurable intelligent surfaces (RISs) are able to programmatically manipulate the propagation environment and hence promising for future evolution of wireless communications \cite{DBLP:journals/corr/abs-1912-07794}. An RIS is a planar meta-surface equipped with a large number of low-cost passive reflective elements, each being able to induce a phase and/or amplitude change to the incident signal independently \cite{8910627}. As such, it can provide passive beamforming gain by optimizing the reflection coefficients on each element. Recently, beamforming design in RIS-aided communications has gained tremendous research attention to enhance system throughput and energy efficiency. 

Most of existing works on RIS-aided communications are limited to one-way passive beamforming design, i.e., the RIS reflection coefficients or phase-shift matrices are optimized for either downlink or uplink transmission, but not both simulaneously. For example, the work \cite{8811733} studies the beamforming design problem to minimize the transmit power in the RIS-aided downlink system under the constraint of users' individual signal-to-interference-plus-noise (SINR) ratio. 
Both the work \cite{8855810} and \cite{8982186} study the beamforming design of the RIS-aided downlink systems to maximize the sum rate under the constraint of the transmit power at the base station (BS).
The work \cite{9270033} considers the weighterd sum-power minimization under quality-of-service (QoS) constraints in the multiuser RIS-aided uplink system.
 The work \cite{2019arXiv191200820H} considers the beamforming design in both downlink and uplink transmissions. But since it assumes time-division duplexing (TDD) systems where the downlink and uplink transmissions occur in different time slots, the passive beamforming for downlink and uplink can be treated separately and still belongs to one-way optimization. 

This work aims to investigate the two-way passive beamforming design in RIS-aided frequency division duplexing (FDD) systems. Unlike the TDD system, the FDD system allows downlink and uplink transmission to take place simultaneously, but over different frequency bands. The operating frequency band of an RIS is typically large enough to cover both the downlink and uplink frequency bands. This suggests that the same RIS phase-shift matrix should be used in both communication directions. 

In this paper, we formulate an optimization problem for maximizing the weighted sum rate of the downlink and uplink transmission in a single-user FDD system where the BS has multiple antennas and the user has single antenna. By varying the weighting parameter, we can set different priorities on the downlink and uplink transmissions and hence obtain different achievable downlink-uplink rate pairs. The problem involves the joint optimization of the phase-shift matrix at the RIS and the transmit and receive beamforming vectors at the BS. This problem is non-convex and highly challenging due to the unit-modulus constraints of the RIS phase shifts. To solve this problem, we propose a manifold optimization algorithm to obtain the stationary solution. Note that the work \cite{DBLP:journals/corr/abs-2008-00448} studies a similar bi-directional beamforming problem for sum-rate maximization in an RIS-aided full-duplex (FD) system. Therein, a fast converging alternating algorithm is proposed. However, this algorithm is limited to equal weights in the sum rate and cannot be extended to arbitary weights as in our work. For comparison, we also introduce two heuristic designs based on one-way optimization, namely, time-sharing and phase-averaging. 

Simulation results show that compared with the two heuristic designs and the one-way only optimization designs, the proposed manifold-based two-way design achieves higher weighted sum rate. It is also found that the manifold-based two-way design significantly enlarges the achievable downlink-uplink rate region compared to the two heuristic designs. Furthermore, when the number of RIS element is large, the phase-averaging design outperforms time-sharing in terms of improving the weighted sum rate.

\vspace{-1.5cm}
\section{System Model and Problem Formulation}
\subsection{System Model}
\begin{figure}[t]
\begin{centering}
\vspace{-0.1cm}
\includegraphics[scale=0.26]{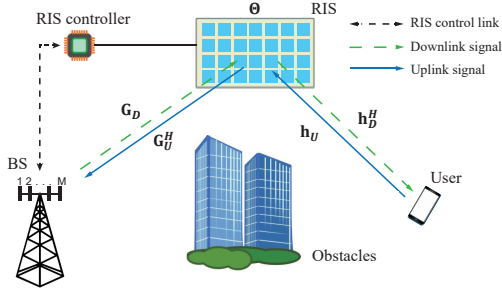}
\vspace{-0.2cm}
 \caption{\small{An RIS-aided FDD communication system}}\label{fig_sys}
\end{centering}
\vspace{-0.4cm}
\end{figure}
As shown in Fig.~\ref{fig_sys}, we consider both downlink and uplink communications in an RIS-aided FDD single-user system, where one BS, equipped with $M$ antennas, communicates with one single-antenna user, via the help of one $F$-element RIS. We assume the direct link between the BS and the user is blocked. Considering the high free-space path loss, we ignore the power of the signals that are responded by the RIS more than one times. In addition, the quasi-static flat-fading model is adopted for all channels and all channel state information (CSI) is perfectly known at a central controller. 

FDD allows the downlink and uplink transmissions at the same time but over different carrier frequency bands. It is also assumed that both the uplink and downlink carrier frequencies are within the operating bandwidth of the RIS. Hence, the same diagonal phase-shift matrix at the RIS should be used for both uplink and downlink transmissions, and can be represented as $\mathbf{\Theta}=\text{diag}(b_1,\ldots,b_F)\in \mathbb{C}^{F\times F}$ with $b_f=e^{j\theta_f}$, where $\theta_f \in [0,2\pi)$ is the phase shift of the $f$-th reflecting element.

Let $s_D \in\mathbb{C}$ and $s_U \in \mathbb{C}$ denote the downlink and uplink information signals, respectively, $n_D \in \mathbb{C}$ and $\mathbf{n}_U \in \mathbb{C}^{M \times 1}$ denote the additive white Gaussian noise (AWGN) at the user with zero mean and power $\sigma_D^2$ and at the BS with $\sigma_U^2$ being the noise power, respectively. Hence, the equivalent received signals at the baseband for the downlink and uplink can be respectively written as:
\begin{subequations}\nonumber
\begin{align}
y_D &=\mathbf{h}_{D}^H\mathbf{\Theta}\mathbf{G}_D\mathbf{w}s_D+n_D,\label{dlreceive} \\
y_U&=\mathbf{v}^H\mathbf{G}_U^H\mathbf{\Theta}\mathbf{h}_{U}\sqrt{P_U}s_U+\mathbf{v}^H\mathbf{n}_U, \label{ulreceive}
\end{align}
\end{subequations}
where $\mathbf{w}\in\mathbb{C}^{M\times 1}$ and $\mathbf{v}\in\mathbb{C}^{M\times1}$ respectively denote the downlink transmit and uplink receive beamforming vectors of the BS, $\mathbf{G}_D \in \mathbb{C}^{F \times M}$ and $\mathbf{G}_U^H \in \mathbb{C}^{M \times F}$ respectively denote the downlink and uplink channel matrices between the RIS and the BS, $\mathbf{h}_{D}^H \in \mathbb{C}^{1\times F}$ and $\mathbf{h}_{U} \in \mathbb{C}^{F \times 1}$ respectively denote the downlink and uplink channel vectors between the RIS and the user, and $P_U \in \mathbb{R}$ denotes the transmit power of the user.

Thus, the SNRs of the downlink and uplink can be written as, respectively:
 \begin{subequations}\nonumber
 \begin{align}\nonumber
 \text{SNR}_D &=\frac{\vert \mathbf{h}_{D}^H\mathbf{\Theta}\mathbf{G}_D\mathbf{w}\vert^2}{\sigma_D^2},\\
\text{SNR}_U &=\frac{\vert\mathbf{v}^H\mathbf{G}_U^H\mathbf{\Theta}\mathbf{h}_{U}\vert^2P_U}{\sigma_U^2}.
\end{align}
\end{subequations}
\subsection{Problem Formulation}
Our objective is to maximize the weighted sum of the downlink rate and uplink rate. The problem is formulated as:
\begin{subequations}
\begin{align}
&{\mathcal{P}_0:}&{\max \limits_{\mathbf{\Theta},\mathbf{w},\mathbf{v}}}\quad & {\eta}\log(1+\text{SNR}_D)+(1-\eta)\log(1+\text{SNR}_U) \label{objf}\\
&{}&{\text{s.t.}}\quad &\left\|\mathbf{w}\right\|_2^2 \leq P_{D,\text{max}}, \label{01}\\
&{}&{}&{P_U} \leq P_{U,\text{max}}, \label{02}\\
&{}&{}&{\vert b_f\vert}=1,\quad \forall f\in\mathcal F, \label{03}
\end{align}
\end{subequations}
where $\eta \in [0,1]$ is a weighting parameter between the downlink rate and the uplink rate, $\mathcal F\triangleq\{1,2,\ldots,F\}$ denotes the index set of reflecting elements of the RIS, $P_{D,\text{max}}$ and $P_{U,\text{max}}$ denote the peak power of the downlink transmission and uplink transmission, respectively. In this problem formulation, by varying the weighted parameter $\eta$, we can obtain an achievable downlink-uplink rate region.
\vspace{-0.2cm}
\section{Algorithm Design}
In this section, we propose a manifold-based algorithm to solve problem $\mathcal{P}_0$. Specifically, we first obtain closed-form solutions for the optimal $\mathbf{w}$ and $\mathbf{v}$ when $\mathbf{\Theta}$ is given. Then, based on these solutions, the original problem $\mathcal{P}_0$ can be converted to a problem which only requires to optimize $\mathbf{\Theta}$.  This problem is then efficiently solved by using manifold optimization.
\subsection{Optimal Transmit and Receiver Beamforming Vectors at BS}
When $\mathbf{\Theta}$ is fixed, problem $\mathcal{P}_0$ can be decoupled into two subproblems: the downlink rate maximization and the uplink rate maximization.

Specifically, the downlink subproblem is
\begin{align}\nonumber
&{} &{\max \limits_{\mathbf{w}:\left\|\mathbf{w}\right\|_2^2\leq P_{D,\text{max}}}} \quad & \log(1+\text{SNR}_D).
\end{align}
It is not difficult to observe that the optimal downlink beamforming is the well-known maximum ratio transmission:
\begin{equation}\label{w}
\mathbf{w}^{*}=\frac{\sqrt{P_{D,\text{max}}}\tilde{\mathbf{h}}_D^H}{\|\tilde{\mathbf{h}}_D\|_2},
\end{equation}
where $\tilde{\mathbf{h}}_D=\mathbf{h}_{D}^H\mathbf{\Theta}\mathbf{G}_D \in \mathbb{C}^{1\times M}$ and it represents the RIS-aided downlink channel vector from the BS to the user.

Similarly, the uplink subproblem is
\begin{align}\nonumber
&{} &{\max \limits_{\mathbf{v},P_U\leq P_{U,\text{max}}}} \quad & \log(1+\text{SNR}_U).
\end{align}
Obviously, the optimal uplink power is $P_U^{*}=P_{U,\text{max}}$ and the optimal uplink receive beamforming is the maximum ratio combining receiver
\begin{equation}\label{v}
\mathbf{v}^{*}=\frac{\tilde{\mathbf{h}}_U^H}{\|\tilde{\mathbf{h}}_U\|_2},
\end{equation}
where $\tilde{\mathbf{h}}_U=\mathbf{h}_{U}^H\mathbf{\Theta}^H\mathbf{G}_U \in\mathbb{C}^{1\times M}$ and it represents the RIS-aided uplink channel vector from the user to the BS.
\vspace{-0.2cm}
\subsection{Manifold Optimization for RIS Phase Shifts}
Denote $\mathbf{b}\triangleq[b_1,\ldots,b_F]^H\in \mathbb{C}^{F \times 1}$, $\mathbf{C}_D\triangleq\text{diag}(\mathbf{h}_{D}^H)\mathbf{G}_D\in\mathbb{C}^{F \times M}$ and $\mathbf{C}_U\triangleq\text{diag}(\mathbf{h}_U^T)\bar{\mathbf{G}}_{U}\in \mathbb{C}^{F\times M}$, where $\bar{\mathbf{G}}_U$ denotes the conjugate of $\mathbf{G}_U$. By substituting the optimal expressions \eqref{w} and \eqref{v} into the objective function \eqref{objf}, problem $\mathcal{P}_0$ is reduced to the following problem:
\begin{subequations}
\begin{align}
&{\mathcal{P}_{1:}}&{\max \limits_{\mathbf{b}}}\quad & f(\mathbf{b})={\eta}f_1(\mathbf{b})+(1-\eta)f_2(\mathbf{b}) \label{8a}\\
&{}&{\text{s.t.}}\quad & {\vert b_f\vert}=1,\quad \forall f\in\mathcal F, \label{8b}
\end{align}
\end{subequations}
where
\begin{equation}
\begin{split}
&f_1(\mathbf{b})=\log\biggl(1+\frac{P_{D,{\text{max}}}\left\|\mathbf{b}^H\mathbf{C}_D\right\|^2}{\sigma_D^2}\biggr), \\
&f_2(\mathbf{b})=\log\biggl(1+\frac{P_{U,\text{max}}\left\|\mathbf{b}^H\mathbf{C}_U\right\|^2}{\sigma_U^2}\biggr).\\
\end{split}
\end{equation}
Note that both the objective function in \eqref{8a} and constraints in \eqref{8b} are non-convex. Also note that $f(\mathbf{b})$ is continuous and differentiable and constraints \eqref{8b} can form a complex circle manifold $\mathcal{M}_{cc}^F=\{\mathbf{b}\in\mathbb{C}^F:\vert b_1\vert=\vert b_2 \vert=\ldots=\vert b_F\vert=1\}$  \cite{7397861}, \cite{AbsMahSep2008}. Therefore, we can adopt the \textit{Riemannian conjugate gradient} (RCG) algorithm to obtain the stationary solution of $\mathcal{P}_1$ \cite{AbsMahSep2008}. The RCG algorithm to update $\mathbf{b}$ is outlined in $\textit{Algorithm}$ $\ref{al1}$.


In the following, we briefly review some basic concepts in manifold optimization and the general produre of the RCG algorithm. More details about manifolds can be found in\cite{AbsMahSep2008}.

A \textit{manifold} $\mathcal{M}$ is a topological space where each point has a neighborhood that is homeomorphic to the Euclidean space. The \textit{tangent space} of the manifold $\mathcal{M}$ at a given point $x\in\mathcal{M}$, denoted as $T_x\mathcal{M}$, is composed of the tangent vectors of all smooth curves through the point $x$. The \textit{Riemannian manifold} is a special kind of manifolds equipped with Riemannian metric. The Riemannian metric allows one to measure distances and use calculas on manifold. This makes it possible to define gradients of cost functions and conduct gradient descent on the Riemannian manifold.

The key idea of the RCG algorithm is to construct a set of conjugate directions by using the gradient at a given point, and search along this set of directions to find the minimum point of the objective function. All these operations are conducted on the Riemannian manifolds.
 The RCG-based $\textit{Algorithm}$ $\ref{al1}$ for solving $\mathcal{P}_1$ consists of the following key steps.

\subsubsection{Find \textit{Riemannian gradient}}The \textit{Riemannian gradient} represents the direction of the greatest increase of a function. The Riemannian gradient of a function at a given point in the manifold is the orthogonal project of its Euclidean gradient onto the tangent space. More specifically, the Riemaninian gradient of the objective function (\ref{8a}), denoted as $\text{grad}f(\mathbf{b})$ at $\mathbf{b}\in\mathcal{M}_{cc}^F$ can be obtained as
\begin{equation}\nonumber
\mathrm{grad}f(\mathbf{b})=\text{Proj}_{\mathbf{b}}\nabla f(\mathbf{b})\\
=\nabla f(\mathbf{b})-\Re\{\nabla f(\mathbf{b})\circ \mathbf{\bar{b}}\}\circ \mathbf{b}\label{gr},
\end{equation}
where the Euclidean gradient of $f(\mathbf{b})$ is
\begin{align}
\begin{split}\label{egr}
\nabla f(\mathbf{b})=\frac{\eta P_{D,\text{max}}\mathbf{C}_D\mathbf{C}_D^H\mathbf{b}}{\sigma_D^2\biggl(1+\frac{P_{D,\text{max}}\left\|\mathbf{b}^H\mathbf{C}_D\right\|^2}{\sigma_D^2}\biggr)}\\
+\frac{(1-\eta)P_{U,\text{max}}\mathbf{C}_U\mathbf{C}_U^H\mathbf{b}}{\sigma_U^2\biggl(1+\frac{P_{U,\text{max}}\left\|\mathbf{b}^H\mathbf{C}_U\right\|^2}{\sigma_U^2}\biggr)}.
\end{split}
\end{align}

\subsubsection{\textit{Retraction}}When a point moves along a tangent vector, it may not be still on the manifold. The Retraction operation can map the point back to the manifold. This operation is used in Step 4 of $\textit{Algorithm}$ \ref{al1} to retract a vector along the Riemannian gradient $\mathbf{d}$ at point $\mathbf{b}_k$ back to the complex circle manifold. This can be stated as:
\begin{align} R_{\mathbf{b}}(\alpha\mathbf{d})=\text{vec}\biggl[\frac{(\mathbf{b}+\alpha\mathbf{d})_i}{\vert(\mathbf{b}+\alpha\mathbf{d})\vert}_{i}\biggr],
\end{align}
where $\alpha$ denotes the step size in the current iteration obtained by Armijo backtracking line search to guarantee the objective function (\ref{8a}) to be non-decreasing \cite{armijo1966}.

\subsubsection{\textit{Transport}}Since the operations between two vectors from different tangent spaces cannot be conducted directly, transport is needed to map one vector from one tangent space to the other tangent space. This operation is used in Step 6 of $\textit{Algorithm}$ \ref{al1} to transport the updated Riemannian gradient $\mathbf{g}_k$. The transport of a tangent vector $\mathbf{d}$ from the tangent space $T_{\mathbf{b}_k}\mathcal{M}_{cc}^F$ at point $\mathbf{b}_k$ to the tangent space $T_{\mathbf{b}_{k+1}}\mathcal{M}_{cc}^F$ at point $\mathbf{b}_{k+1}$ can be stated as:
\begin{align}\nonumber
\text{Transp}_{{\mathbf{b}}_k \rightarrow {\mathbf{b}_{k+1}}}: & {T}_{\mathbf{b}_{k}}\mathcal{M}_{cc}^F\rightarrow T_{\mathbf{b}_{k+1}}\mathcal{M}_{cc}^F:\\
{}&{\mathbf{d}}\rightarrow \mathbf{d}-\Re\{\mathbf{d}\circ \bar{\mathbf{b}}_{k+1}\}\circ{\mathbf{b}_{k+1}}.
\end{align}

\begin{algorithm}[t]
	\caption{RCG Algorithm for RIS Phase Shifts}
	\label{al1}
	\hspace*{0.02in} {\bf Input:} $\eta$, $\mathbf{C}_D$, $\mathbf{C}_U$, $\sigma_D^2$, $\sigma_U^2$
	\begin{algorithmic}[1]
		\State Initialize $\mathbf{b}_0\in\mathcal{M}_{cc}^F$, $\mathbf{d}_0=\mathrm{grad}f(\mathbf{b}_0)$, and set the iteration number $k=0$;
		\Repeat
		\State Choose Armijo backtracking line search step size $\alpha_k$;
		\State Find the next point $\mathbf{b}_{k+1}\leftarrow R_{\mathbf{b}_k}(\alpha_k\mathbf{d}_k)$;
		\State Update Riemannian gradient $\mathbf{g}_{k+1}=\mathrm{grad}f(\mathbf{b}_{k+1})$ according to $\eqref{gr}$;
		\State Calculate the vector transports $\mathbf{g}_k^{+}$ and $\mathbf{d}_k^{+}$ of gradient $\mathbf{g}_{k}$ and conjugate direction $\mathbf{d}_k$ from $\mathbf{b}_k$ to $\mathbf{b}_{k+1}$;
		\State Choose Polak-Ribiere parameter $\beta_{k+1}$;
		\State Compute conjugate direction $\mathbf{d}_{k+1}=\mathbf{g}_{k+1}+\beta_{k+1}\mathbf{d}_k^{+}$;
		\State Update $k=k+1$
		\Until{a stopping criterion is met.}
	\end{algorithmic}
\end{algorithm}

 In Step 7 and 8 of $\textit{Algorithm}$ \ref{al1}, the conjugate direction is constructed. The Polak-Ribiere parameter $\beta_{k+1}$ can be obtained by using the method in Chapter 1 of \cite{bertsekas1997nonlinear}. Note that this algorithm can obtain a stationary solution because it will converge to a critical point where the gradient of the objective function is zero.

 Next, we introduce the complexity of $\textit{Algorithm}$ \ref{al1}. The complexity of obtaining optimal transmit and receiver beamforming vectors at the BS is $\mathcal{O}(F^2+FM)$. Besides, the complexity of the RCG algorithm is dominated by computing the Euclidean gradient, which is $\mathcal{O}(FM+F^2M+F^2)$. The retraction step also requires iteratively searching $\alpha$ and the complexity is $\mathcal{O}(F)$. Therefore, the total complexity of the RCG-based two-way optimization algorithm is $\mathcal{O}(I_R(F+FM+F^2M+F^2)+F^2+FM)$, where $I_R$ denotes the iteration times of the RCG algorithm.
\vspace{-0.1cm}
\section{Heuristic Designs of RIS Phase Shifts}
The RCG-based algorithm proposed in the previous section is an iterative algorithm aiming for two-way optimization. Considering the implementation complexity, in this section, we propose two heuristic low-complexity designs for the RIS phase shift based on traditional one-way optimization. The purpose is to illustrate the advantage of two-way optimization over one-way design. We first obtain the optimal one-way RIS response matrices for downlink and uplink, separately. We then introduce the two heuristic methods to find the two-way RIS response matrix based on time sharing and phase averaging, respectively. 
\subsection{One-Way RIS Phase Shift Design}
We adopt the alternating optimization algorithm to alternatively optimize beamforming vectors and RIS phase shifts in the downlink and uplink parts, respectively. 

For a given $\mathbf{w}$, the downlink rate maximization problem can be represented as
\begin{subequations}
\begin{align}
&{\mathcal{P}_{2-1}:}&{\max \limits_{\mathbf{b}}}\quad & \log\biggl(1+\frac{\vert\mathbf{b}^H\mathbf{J}_D\vert^2}{\sigma_D^2}\biggr)\\
&{}&{\text{s.t.}}\quad & {\vert b_f\vert}=1,\quad \forall f\in\mathcal F,
\end{align}
\end{subequations}
where $\mathbf{J}_D\triangleq \text{diag}(\mathbf{h}^H)\mathbf{G}_D\mathbf{w}\in\mathbb{C}^{F \times 1}$.
Obviously, the optimal solution of problem $\mathcal{P}_{2-1}$ is given by:
\begin{equation}
\mathbf{b}_D^*=e^{j\text{arg}\{\mathbf{J}_D\}} \label{bD}.
\end{equation}

Similarly, for a given $\mathbf{v}$, the uplink rate maximization problem can be represented as
\begin{subequations}
\begin{align}
&{\mathcal{P}_{2-2}:}&{\max \limits_{\mathbf{b}}}\quad & \log\biggl(1+\frac{P_{U,\text{max}}\vert
\mathbf{b}^H\mathbf{J}_U\vert^2}{\sigma_U^2}\biggr)\\
&{}&{\text{s.t.}}\quad & {\vert b_f\vert}=1,\quad \forall f\in\mathcal F,
\end{align}
\end{subequations}
where $\mathbf{J}_U\triangleq \text{diag}(\mathbf{v}^H\mathbf{G}_U^H)\mathbf{h}_U \in \mathbb{C}^{F \times 1}$. The optimal solution of problem $\mathcal{P}_{2-2}$ is given by:
\begin{equation}\nonumber
\mathbf{b}_U^*=e^{j\text{arg}\{\mathbf{J}_U\}}\label{bU}.
\end{equation}
\subsection{Time-Sharing Method}
Recall that the objective of the problem $\mathcal{P}_0$ is to maximize the weighted sum-rate of the downlink and uplink with weighting parameter given by $\eta\in [0,1]$. In the extreme case when $\eta=1$ (or $0$), the objective is to maximize the downlink rate (or uplink rate) only. As such, the first heuristic design of $\mathbf{b}$ is to apply time-share between the two one-way solutions introduced above, i.e., for a fraction of time, $\eta$, the system applies the optimal downlink $\mathbf{b}_D$, and for the rest fraction of time, $1-\eta$, the system applies the optimal uplink $\mathbf{b}_U$. Let $r_D$ and $r_U$ denote the ultimate achievable downlink and uplink rate of this design, respectively. In specific, $r_D$ and $r_U$ are given by:
\begin{align}
\begin{split}\label{t}
r_D &=\eta \log(1+\text{SNR}_D(\mathbf{b}_D^*))+(1-\eta)\log(1+\text{SNR}_D(\mathbf{b}_U^*)) \\
r_U &=\eta \log(1+\text{SNR}_U(\mathbf{b}_D^*))+(1-\eta)\log(1+\text{SNR}_U(\mathbf{b}_U^*)).
\end{split}
\end{align}
\vspace{-0.5cm}
\subsection{Phase-Averaging Method}
Similarly, based on the weighting parameter in the objective of $\mathcal{P}_0$, the second heuristic design of $\mathbf{b}$ is to apply phase-averaging between the two one-way solutions on each individual element with $\eta$ and $(1-\eta)$ being the corresponding weights for averaging. In specific, the RIS phase shift $\mathbf{b}$ is given by:
\begin{equation}\nonumber
\mathbf{b}=e^{j\{\eta\text{arg}{(\mathbf{b}_D^*)}+(1-\eta)\text{arg}{(\mathbf{b}_U^*)}\}}.
\end{equation}

These two heuristic designs are computationally efficient as all variables are updated by using closed-form expressions. To be specific, it can be shown that the complexity of updating $\mathbf{w}$ (or $\mathbf{v}$) is $\mathcal{O}(F^2+FM)$, that of obtaining $\mathbf {b}_D^*$ is $\mathcal{O}(FM)$, and that of obtaining $\mathbf{b}_U^*$ is $\mathcal{O}(FM+F^2)$. Thus, the complexity of two heuristic designs is $\mathcal{O}((I_D+I_U)(F^2+FM))$, where $I_D$ and $I_U$ denote the iteration times of the downlink RIS phase shift design and the uplink phase shift design, respectively. Simulation results will show that both $I_D$ and $I_U$ are generally much smaller than $I_R$. Hence, the complexity of these two heuristic designs is lower than the manifold-based two-way optimization algorithm.
\section{Simulation Results}
In this section, we consider a uniform linear array (ULA) structure at the BS with $M=4$ antennas located at ($0$m, $0$m). The RIS has a uniform planar array (UPA) structure located at ($d$m, $5$m) and equipped with $F_1\times F_2$ reflecting elements where $F_1=10$ and $F_2$ can vary. The user is located at ($50$m, $0$m). The antenna spacing is half of the downlink wavelength.

We assume the Rician fading channel model for all channels involved. Thus, the BS-RIS channel is given by:
\begin{equation}\nonumber
\mathbf{G} = \sqrt{L(d,f)}\left( \sqrt{\frac{\beta_{BR}}{1+\beta_{BR}}}\mathbf{G}^{\rm LoS} + \sqrt{\frac{1}{1+\beta_{BR}}}\mathbf{G}^{\rm NLoS} \right),
\end{equation}
where $\beta_{BR}$ is the Rician factor, $\mathbf{G}^{\rm LoS}$ and $\mathbf{G}^{\rm NLoS}$ are line-of-sight(LoS) and non-line-of-sight(NLoS) components, and $L(d,f)$ denotes the corresponding path loss.

The LoS component is given by:
\begin{equation}\nonumber
\mathbf{G}^{\mathrm{LoS}} = \mathbf{a}_r(\vartheta) e^{-j\phi} \mathbf{a}_t(\psi),
\end{equation}
where $\mathbf{a}_t$ and $\mathbf{a}_r$ are steering vectors, $\vartheta$ and $\psi$ are the angular parameters and $\phi$ is the phase difference on the propagation path. The elements of NLoS component are chosen from $\mathcal{CN}(0,1)$. The RIS-user channel is also generated by following the similar procedure. We set Rician factors of BS-RIS channel and RIS-user channel as $\beta_{\rm BR}=2$ and $\beta_{\rm Ru}=0.5$, respectively.

According to the 3GPP specification \cite{3gpp.36.814}     and other existing work \cite{8811733}, the path loss is frequency-dependent and distance-dependent, which is given by:
\begin{equation}\nonumber
L(d,f) = C_0\left( \frac{f}{f_0} \right)^{-2} \left( \frac{d}{D_0} \right)^{-\alpha} ,
\end{equation}
where $C_0$ is the path loss at the reference distance $D_0=1$ meter (m) and the reference frequency $f_0=1$ GHz, $f$ denotes the center frequency of the signal, $d$ denotes the link distance, and $\alpha$ is the path loss exponent. We set the path loss exponents of BS-RIS channel and RIS-user channel as $\alpha_{\rm BR} = 2$ and $\alpha_{\rm Ru} = 2.8$, respectively.

If not specified otherwise, we set $F_2 = 6$, $P_{U,\rm max} = 0.5$ W, $P_{D,\rm max} = 5$ W, $\sigma_U^2 = -70$ dBm, $\sigma_D^2 = -70$ dBm, $C_0 = -30$ dB.  The downlink and uplink carrier frequencies are $f_D = 1855$ MHz and $f_U = 1760$ MHz, used in China Unicom LTE FDD .
\begin{figure}[t]
	\begin{centering}
		\includegraphics[scale=0.42]{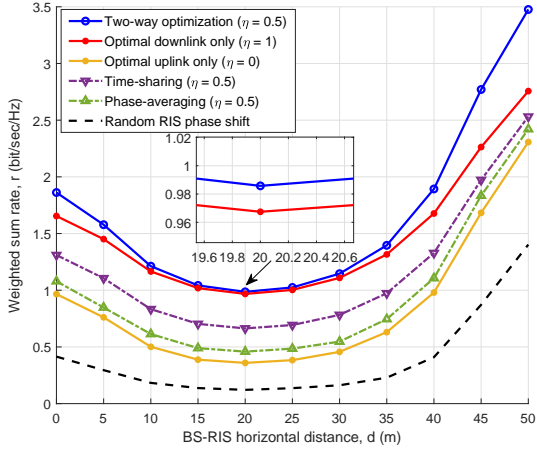}
		\caption{\small{Weighted sum rate ($\eta=0.5$) versus BS-RIS horizontal distance}}\label{fig_dis1}
	\end{centering}
	\vspace{-0.4cm}
\end{figure}
\subsection{Weighted Sum Rate Versus BS-RIS Horizontal Distance}
In Fig.~\ref{fig_dis1}, we compare the weighted sum rate obtained by different schemes at $\eta=0.5$ when the horizontal distance between the BS and the RIS, $d$, varies. First, it is observed that the proposed two-way optimization scheme which jointly optimizes the downlink and uplink transmissions outperforms other benchmark schemes.  Second,  for all schemes, the RIS near either the BS (e.g., $d=5$ m) or the user (e.g., $d=45$ m) achieves higher weighted sum rate than the RIS in the middle of them (e.g., $d=20$ m). In addition, compared with the case where the RIS near the BS (e.g., $d=3$ m), higher weighted sum rate can be achieved by the case where the RIS near the user (e.g., $d=47$ m).

It is also observed from Fig.~\ref{fig_dis1} that the scheme which only optimizes the downlink transmission performs close to the two-way optimization scheme when the RIS is in the middle of the BS and the user (e.g., $d=20$ m), while it achieves considerably lower weighted sum rate when the RIS is nearer to the user (e.g., $d=50$ m). This is expected since in the former case, the downlink transmission allocated much higher transmit power than the uplink transmission.

As for the two heuristic designs, it is found that the time-sharing method outperforms the phase-averaging method at the given simulation setting. In addition, they both outperform the optimal uplink only scheme.
\begin{figure}[t]
	\begin{centering}
		\includegraphics[scale=0.42]{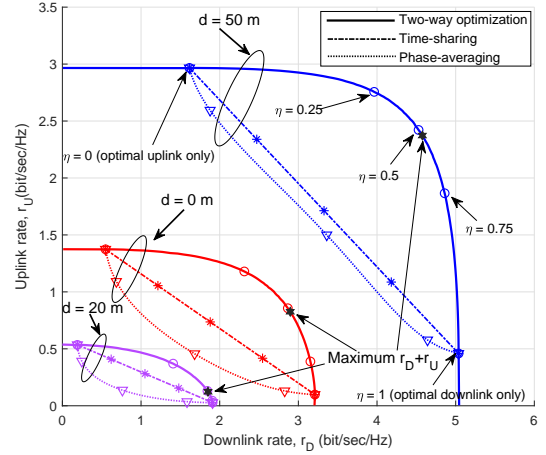}
		\caption{\small{The downlink and uplink rate region }}\label{fig_reg1}
	\end{centering}
	\vspace{-0.1cm}
\end{figure}
\begin{figure}[t]
	\begin{centering}
		\includegraphics[scale=0.40]{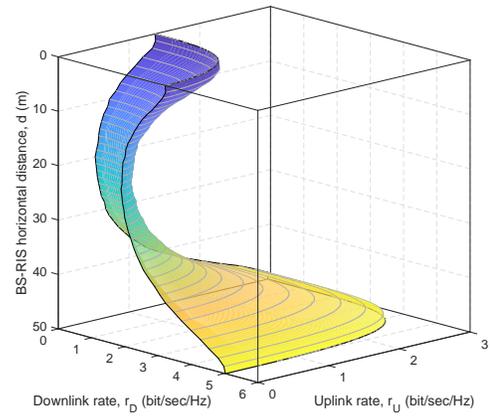}
		\caption{\small{Achievable downlink-uplink rate of the two-way optimization scheme at different BS-RIS horizontal distance}}\label{fig_reg2}
	\end{centering}
	\vspace{-0.4cm}
\end{figure}
\subsection{Rate-Region Comparison}
The comparison of the achievable downlink-uplink rate region at $d=50$m, $20$m, and $0$m is illustrated in Fig.~\ref{fig_reg1}. In the two-way optimization and phase-averaging schemes, the achievable downlink and uplink rates are defined as $r_D\triangleq\log(1+\text{SNR}_D)$ and $r_U\triangleq \log(1+\text{SNR}_U)$, respectively. In the time-sharing scheme, the achievable downlink and uplink rates are represented as \eqref{t}, respectively. For the two-way optimization scheme, the downlink-uplink tradeoff curves are obtained by controlling the weighting parameter $\eta$ in $\mathcal{P}_0$. When $\eta=0$ or $1$, the objective only accounts for the downlink transmission rate or the uplink transmission rate. It is obvious that the achievable downlink-uplink transmission rate region of the two-way optimization scheme is much larger than that of the time-sharing scheme. In specific, when $d=50$ m and $\eta=0.5$ (in the time-sharing scheme), compared with the time-sharing scheme, the two-way optimization scheme increases the uplink transmission rate from $1.8$ bit/sec/Hz to $2.9$ bit/sec/Hz with the same downlink transmission rate of $3.3$ bit/sec/Hz, which is $61\%$ higher. It also increases the downlink transmission rate from $3.3$ bit/sec/Hz to $4.8$ bit/sec/Hz with the same uplink transmission rate of $1.8$ bit/sec/Hz, resulting in $45\%$ improvement. Moreover, it also shows that the phase-averaging scheme achieves the smallest downlink-uplink transmission rate region. In addition, for all the RIS deployment cases shown in  Fig.~\ref{fig_reg1}, when $\eta$ is around 0.5, the two-way optimization scheme can achieve the maximum achievable sum rate $r_D+r_U$.

More details are shown in Fig.~\ref{fig_reg2}. From this figure, it is clear that among all of the considered RIS deployment cases, the case when RIS is the nearest to the user ($d=50$m) achieves the largest downlink-uplink rate region.
\subsection{Effects of Reflecting Elements}
\begin{figure}[t]
	\begin{centering}
		\includegraphics[scale=0.40]{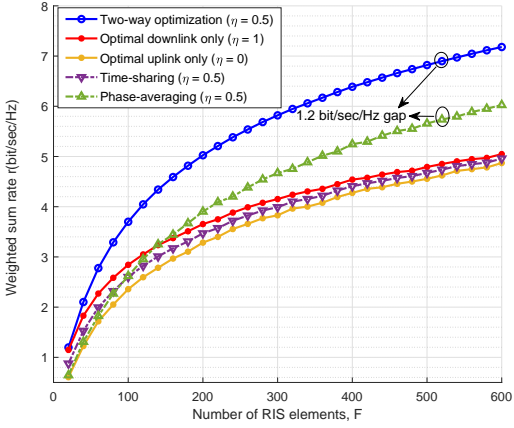}
		\caption{\small{Weighted sum rate versus the number of reflecting elements at the RIS}}\label{fig_ele}
	\end{centering}
	\vspace{-0.4cm}
\end{figure}
In Fig.~\ref{fig_ele}, we compare the weighted sum rate of all the above schemes versus the number of reflecting elements at the RIS when $d=45$ m.
As expected, the two-way optimization scheme achieves the best performance among all the considered schemes for all $F$. The more interesting observation lies at the performance comparison among the two heuristic scheme and the optimal uplink/downlink only schemes. In specific, when the number of reflection elements is small, the optimal-downlink only scheme is better than the two heuristic schemes. But as $F$ increases, the performance of the phase-averaging scheme improves faster than the one-way optimization scheme(i.e., only optimize downlink scheme and only optimize uplink scheme). When $F$ is large, there is a nearly constant 1.2 bit/sec/Hz gap between the two-way optimization scheme and the phase-averaging scheme. According to the simulation, the computation time of the phase-averaging scheme is only 12.6\% of the two-way optimization scheme.
\section{Conclusion}
This paper investigated the joint design of the transmitting/receiving beamformers at the BS and the RIS phase shifts for an RIS-aided FDD communication system. We formulated an optimization problem to maximize the weighted sum of the downlink rate and the uplink rate subject to peak power constraints. The formulated non-convex problem is optimally solved by using our proposed manifold-based two-way optimization algorithm. To illustrate the advantage of two-way optimization, we also proposed two heuristic methods and they can simplify the procedure of solving the problem. Simulation results demonstrated that our proposed two-way optimization scheme outperforms one-way optimization schemes and two heuristic designs in terms of the weighted sum rate. It is also found that the proposed two-way optimization scheme can achieve larger downlink-uplink rate region compared to two heuristic schemes. Although the weighted sum rate can be further improved by adopting our proposed algorithms in the RIS-aided single-user system, the more practical cases should be further considered in the future work, e.g. the discrete phase shifts problem of the RIS and the multiuser case in FDD system.
\bibliographystyle{IEEEtran}
\bibliography{IEEEabrv,isit15}
\end{document}